\documentclass[aps,prl,twocolumn,showpacs,floats,graphicx,letterpaper,superscriptaddress]{revtex4-1}

\usepackage{graphicx}
\begin{document}


\title{Band-dependent Quasiparticle Dynamics in Single Crystals of the Ba$_{0.6}$K$_{0.4}$Fe$_2$As$_2$ Superconductor Revealed by Pump-Probe Spectroscopy}


\author{Darius H Torchinsky}
\affiliation{Department of Physics, Massachusetts Institute of
Technology, Cambridge, Massachusetts, 02139, USA}

\author{G.F. Chen}
\author{J.L. Luo}
\author{N. L. Wang}
\affiliation{Beijing National Laboratory for Condensed Matter
Physics, Institute of Physics, Chinese Academy of Sciences,
Beijing 100190, China}

\author{Nuh Gedik}%
\email{gedik@mit.edu}\affiliation{Department of Physics,
Massachusetts Institute of Technology, Cambridge, Massachusetts,
02139, USA}

\date{\today}

\begin{abstract}
We report on band-dependent quasiparticle dynamics in ${\rm
Ba_{0.6}K_{0.4}Fe_2As_2}$ (${\rm T_c~=~37~K}$) measured using
ultrafast pump-probe spectroscopy. In the superconducting state,
we observe two distinct relaxation processes: a fast component
whose decay rate increases linearly with excitation density and a
slow component with an excitation density independent decay rate.
We argue that these two components reflect the recombination of
quasiparticles in the two hole bands through intraband and
interband processes. We also find that the thermal recombination
rate of quasiparticles increases quadratically with temperature.
The temperature and excitation density dependence of the decays
indicates fully gapped hole bands and nodal or very anisotropic
electron bands.

\end{abstract}

\maketitle

One of the main factors complicating studies of the iron pnictides
is the existence of multiple electron and hole bands crossing the
Fermi surface \cite{ding2,mazin,chubukov}, each of which develops
a gap upon cooling below the superconducting transition
temperature $T_c$. Consequently, there is still no consensus
within the community concerning the symmetry and structure of the
superconducting order parameter. Angle resolved photoemission
spectroscopy (ARPES) measurements consistently show nearly
isotropic gaps with no evidence of nodes \cite{ding2}, although
the surface-selective nature of ARPES, and its diminished
resolution at the Brillouin zone corner may obscure the presence
of nodes. Meanwhile, many bulk measurements which integrate across
the Brillouin zone have painted a somewhat different picture;
nuclear magnetic resonance (NMR) spin relaxation measurements of
the $1/T_1$ relaxation rate \cite{fukazawa} and penetration depth
experiments \cite{martin} both yield a power law temperature
dependence usually associated with nodes.

Theoretical approaches also suggest a number of different pairing
symmetries both with and without nodes. Mazin et al. proposed a
nodeless extended s-wave gap that changes sign between electron
and hole pockets ($s_{\pm}$) \cite{mazin}, Graser et al. found
$s_{\pm}$ pairing symmetry with nodes on electron pockets or
$d_{x^2-y^2}$ with nodes on hole pockets, depending on the doping
\cite{graser} while Yanagi and coworkers proposed a $d_{xy}$ state
with nodes on both the electron and hole pockets \cite{yanagi}.
These different scenarios arise as a result of the interplay
between intra and interband interactions \cite{chubukov2}. It is
therefore of crucial importance to be able to experimentally probe
the low energy excitations of these materials in a band-sensitive
manner in order to better elucidate the superconducting order
parameter.

Time resolved optical pump-probe spectroscopy
\cite{Averitt,segre,gedik,gedik2,mertelj,chia} is a powerful,
bulk-sensitive technique which provides a unique window upon
quasiparticle (QP) interactions in the iron pnictides. In these
experiments, an ultrashort laser pulse is split into two portions:
one of the pulses (pump) is used to inject nonequilibrium
excitations and the other pulse (probe) is used to probe the
subsequent temporal evolution of the density of these excitations
via measurement of the change in reflectivity of the sample,
assumed proportional to the QP population $n$, as a function of
time. These experiments have been performed in the cuprates
\cite{gedik,gedik2,demsar} to study fundamental issues of gap
symmetry and recombination dynamics, and have recently been
applied to the pnictides to provide evidence for a psuedogap state
\cite{mertelj} and the existence of competing electronic order
\cite{chia}. However, ultrafast studies of QP recombination
dynamics which can reveal the structure of the superconducting
energy gap in momentum space have not been previously
investigated.

In this Letter, we present an optical pump-probe study of QP
recombination in ${\rm Ba_{0.6}K_{0.4}Fe_2As_2}$ ($T_c=37~K$) that
reveals band-specific dynamics. Below $T_c$, we find that our
signal is comprised of two distinct components which, aided by LDA
calculations \cite{ma} and ARPES measurements
\cite{ding,ding2,richard,wray}, we assign to the three Fermi
surface hole pockets. The first component, which we assign to the
innermost hole pockets, is characterized by a fast QP decay rate,
which depends linearly on the excitation density, signifying
pairwise recombination. By studying the excitation density
dependence of this decay, we obtained the thermal recombination
rate of QPs, which was found to vary as $\sim T^2$. The second
component is a slow, excitation density independent decay
indicative of ``bottlenecked'' recombination of QPs from a
different component of the superconducting system, which we assign
to the outer hole band. The number of photoinduced QPs ($n$)
contributing to the overall signal increases linearly with laser
fluence, indicating the fully gapped character of these bands.
These observations are consistent with models which predict fully
gapped hole bands and nodal or very anisotropic gaps in the
electron pockets.

For this study, we used a Ti:sapphire oscillator producing pulses
with center wavelength 795~nm ($h\nu={\rm 1.56 ~eV}$) and duration
60~fs at FWHM. The 80~MHz repetition rate was reduced to 1.6~MHz
with a pulse picker to eliminate steady state heating of the
sample. Both beams were focussed to 60~$\mu$m FWHM spots and the
probe beam reflected back to a photodiode for detection. Use of a
double-modulation scheme \cite{gedik} provided sensitivity of the
fractional change in reflectivity of $\Delta R/R\sim 10^{-7}$. The
pump fluence $\Phi$ was varied with neutral density filters in
order to tune the QP excitation density. High-quality single
crystals of ${\rm Ba_{0.6}K_{0.4}Fe_2As_2}$ were grown by the
self-flux method \cite{chen}. SQUID magnetometry measurements
yielded a very sharp transition ($\Delta T \approx$ 1~K) at ${\rm
T_c~=~37}$~K indicating a high degree of sample purity. Further
characterization of sample quality may be found in the online
supplementary information.

\begin{figure}
\includegraphics[scale=.45]{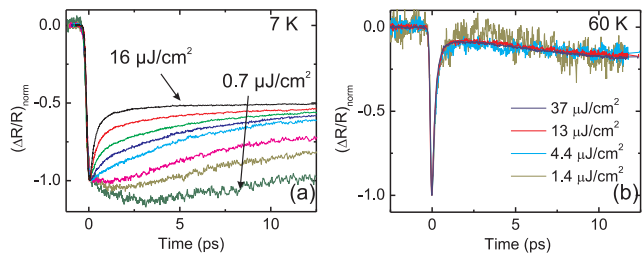}
\caption[Raw data]{\label{fig:rawdata}Fast, intensity-dependent
relaxation of the normalized reflectivity transients $(\Delta
R/R)_{norm}$ in the superconducting state. The decay rate of
$(\Delta R(t)/R)_{norm}$ systematically decreases with decreasing
intensity at 7~K (a) [from bottom to top: $\Phi=$ 0.7~${\rm \mu
J/cm^2}$ (light green), 2.2, 4.4, 7.0, 8.9, 12.8 to 16.1~${\rm \mu
J/cm^2}$ (black)] but not at 60~K (b).}
\end{figure}

Figures~\ref{fig:rawdata}(a) and ~\ref{fig:rawdata}(b) show raw
data traces of $\Delta R(t)/R$ normalized to their $t=0$ value at
7~K and 60~K, respectively, for a variety of $\Phi$. At 7~K,
photoexcitation leads to a decrease in the reflectivity and the
rate of recovery increases with increasing pump fluence. At 60~K,
we observe data collapse of the normalized traces indicating that
all $T>T_c$ recovery dynamics are independent of $\Phi$. From the
data in Figs.~\ref{fig:rawdata}(a) and ~\ref{fig:rawdata}(b), we
conclude that the decay rate at short times increases both with
increasing temperature and excitation density below $T_c$. In
order to rule out steady state heating as the source of this
intensity dependence, we used a pulse picker to vary the
repetition rate from 1.6~MHz to 200~kHz, which produced no
discernible change in the recovery dynamics.

\begin{figure}
\includegraphics[scale=0.45]{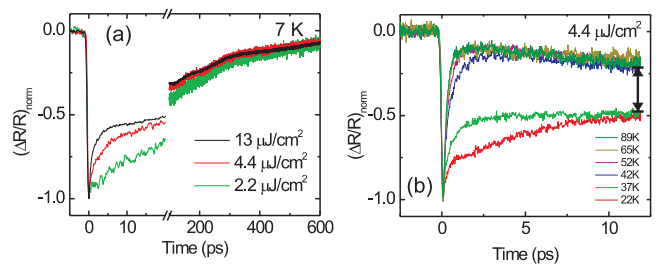}
\caption[Slow component]{\label{fig:slow}Slow relaxation of
$(\Delta R/R)_{norm}$ in the superconducting state. (a) $(\Delta
R/R)_{norm}$ at 7~K measured at three different fluences show no
intensity dependence in the slow dynamics despite their marked
intensity dependence at short times. (b) Temperature dependence of
$(\Delta R/R)_{norm}$ near $T_c$ obtained at $\Phi=4.4~J/cm^2$. We
note a sharp decrease in the offset across the transition (i.e.,
37 to 42~K) due to switching off of the long-time component with
loss of superconductivity. Above $T_c$ there is an upturn of the
signal evident at short times due to stimulated Brillouin
scattering \cite{thomsen}.}
\end{figure}

Closer examination of the data in Fig.~\ref{fig:rawdata}(a)
reveals that after the initial intensity-dependent relaxation,
$\Delta R(t)/R$ tends to a constant offset of $(\Delta
R/R)_{norm}\simeq 0.5$. This offset is the beginning a slow
component in the signal, shown in Fig.~\ref{fig:slow}(a), whose
decay was observed to be intensity independent. As with the fast
intensity dependence, the slow component was observed to switch
off abruptly at $T_c$, signifying its origin in superconductivity
[Fig.~\ref{fig:slow}(b)].

\begin{figure}
\includegraphics[scale=0.4]{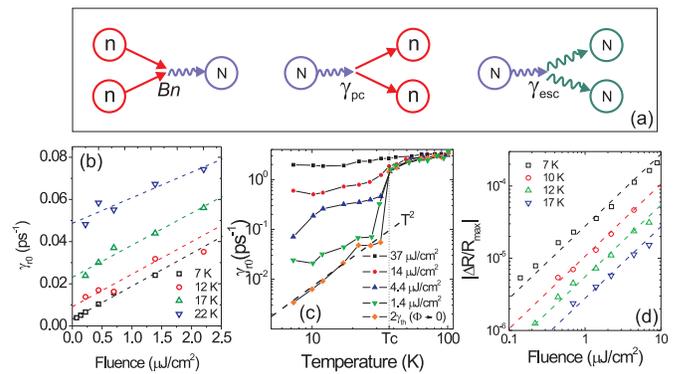}
\caption[Raw data]{\label{fig:Medusa}Decay-rate analysis of data
with a view towards gap symmetry. (a) Illustration of the three
rates in the Rothwarf-Taylor model: Bare QP recombination rate
($Bn$), pair creation rate ($\gamma_{pc}$) and escape rate
($\gamma_{esc}$). ($n$: QP, $N$: Bosons) (b) The initial decay
rate ($\gamma_{r0}$) as a function of pump fluence at various
temperatures. The linear dependence of the recombination rate on
initial population is a hallmark of bimolecular recombination.
Dashed lines show linear fits to data. The thermal decay rate at
each temperature was obtained by extrapolating the fits to zero
fluence. (c) The initial relaxation rate $\gamma_{r0}$ plotted as
a function of temperature for five representative pump fluences
along with the thermal rate $\gamma_{th}$ (orange diamonds). Below
$T_c$, $\gamma_{r0}$ strongly depends on the excitation density.
There is a sharp transition at 37~K, although a discernible
intensity dependence persists above $T_c$. The dashed line shows
$\sim T^2$ dependence as a guide. (d) $|(\Delta R /R)_{max}|$
plotted as a function of $\Phi$ for four temperatures, each offset
for clarity. Dashed lines show a slope of 1.}
\end{figure}

The characteristics of these two separate relaxation timescales
can be understood by considering the coupled dynamics of gap
energy QPs and bosons within the context of the Rothwarf-Taylor
(RT) model \cite{roth-tay}. In this model, there are three
important rates that determine the behavior of the system
\cite{gedik,kabanov}, each depicted schematically in
Fig.~\ref{fig:Medusa}(a). The first is the bare recombination rate
of QPs in which two QPs form a Cooper pair and emit a $2\Delta$
boson. This rate goes as $\gamma_r=-\frac{1}{n}\frac{dn}{dt}$ and
depends linearly on the population $n$ ($\gamma_r=Bn$, where $B$
is the bimolecular recombination coefficient). The emitted boson
can then either recreate the QP pair with rate $\gamma_{pc}$ or
escape with rate $\gamma_{esc}$. This escape ultimately occurs
either via decay into other, lower energy ($<2\Delta$) bosons or
by propagation out of the probed region. The relative magnitudes
of these three rates determines if the density-dependent bare QP
recombination rate can be observed.

The fast component observed in Fig.~\ref{fig:rawdata}(a) is
analyzed by obtaining the initial slope of the transients
following the peak. This initial decay rate ($\gamma_{r0}$)
exhibits bare QP recombination, as evidenced by its linear
increase with excitation density at low fluences, shown by the
black squares in Fig.~\ref{fig:Medusa}(b). The observation of
bimolecular recombination physically originates from either a very
slow pair creation rate ($\gamma_{pc}/Bn\ll 1$) or a very fast
escape rate such that the pairing boson escapes before it has a
chance to break apart a Cooper pair ($\gamma_{esc}\gg\gamma_{pc}$)
\cite{segre}.

In turn, the slow component of Fig.~\ref{fig:slow}(a) reflects a
separate set of dynamics. While prior studies in the cuprates have
suggested that laser heating \cite{mazin2} or QPs trapped in
in-gap states \cite{thomas} may cause a slow decay, the decay
observed here does not fit within either of these proposed
scenarios. The former is precluded by the sudden disappearance of
the slow component at $T_c$ while the latter is ruled out due to
the consistency of the slow component between different sample
batches, where variability in impurity concentration would cause
marked differences. Rather, we ascribe these dynamics to a
separate QP recombination process governed by a different set of
RT parameters where $\gamma_{esc}$ is the slowest i.e.
$\gamma_{esc} \ll \gamma_{pc}$, $Bn$. Here, one observes the boson
bottleneck in which QPs and bosons quickly come to a
quasiequilibrium \cite{kabanov} and the combined population
finally decays with a slow, intensity independent rate
proportional to $\gamma_{esc}$.

We may thus exploit the distinctive relaxation dynamics of the
fast component to obtain clues about the symmetry and structure of
the superconducting gap \cite{gedik} via the thermal decay rate of
QPs participating in recombination. In the low excitation regime,
the thermal density of QPs ($n_{th}$) becomes comparable with the
photoinduced QP density. A photoinduced QP can then either
recombine with another photoinduced QP or a thermally excited QP,
resulting in an aggregate decay rate $\gamma_{r0}=Bn+2Bn_{th}$.
Figure~\ref{fig:Medusa}(b) shows $\gamma_{r0}$ as a function of
laser fluence ($\Phi$) for four representative temperatures. We
observe that at each temperature, $\gamma_{r0}$ increases linearly
in $\Phi$ with a temperature dependent intercept. This linear
dependence is due to the first term in $\gamma_{r0}$ ($Bn$) and
the intercept is the thermal recombination rate
($2\gamma_{th}=2Bn_{th}$ as $n\propto \Phi \rightarrow 0$). The
slope of the linear fits ($B$) exhibits no strong temperature
dependence while the intercept $2Bn_{th}$ increases with $T$ due
to a larger $n_{th}$.

Figure~\ref{fig:Medusa}(c) presents the temperature dependence of
$\gamma_{th}$ and $\gamma_{r0}$ at various fluences, both in the
high and low fluence regimes. The signature of pairwise
recombination is evident through the strong dependence of
$\gamma_{r0}$ on the excitation density, which diminishes markedly
at $T_c$. Significantly, we note that the thermal decay rate
depends quadratically on temperature ($\gamma_{th}=B n_{th}\propto
T^2$) below $T_c$. In the case of a simple isotropic s-wave gap,
$n_{th}$ is expected to depend on $T$ as
$\sim\exp{\left(-\Delta/k_bT\right)}$, whereas the presence of a
node would give rise to a power law dependence. In particular, a
line node in the gap would lead to the observed $n_{th}\propto
T^2$ due to the linear dependence of the density of states on
energy in the node.

A significant piece of information regarding the possible location
of such a node in momentum space is provided by consideration of
the dependence of the photoinduced QP density ($n$) on the total
energy deposited into the system. As mentioned above, the
amplitude of the measured change in reflectivity is proportional
to $n$, whereas the absorbed laser fluence ($\Phi$) is
proportional to the total energy stored in the QP system. For
fully gapped excitations, $n$ is proportional to the total energy
absorbed, and hence we simply have $\Delta R /R\propto \Phi$
\cite{gedik}. In the presence of a line node, the linear
dependence of the density of states on energy implies that $n$
should vary with the total stored energy in a sublinear fashion
($\Delta R /R\propto \Phi^{2/3}$) \cite{gedik}.
Figure~\ref{fig:Medusa}(d) shows the amplitude of $\Delta R /R$ as
a function of $\Phi$ at low fluence for four temperatures, where a
clear linear dependence is observed at all temperatures,
indicating that the photoinduced QPs all originate from fully
gapped excitations.

\begin{figure}
\includegraphics[scale=0.4]{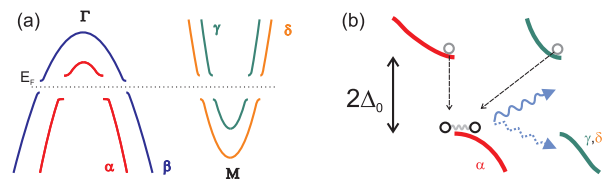}
\caption[RT model]{\label{fig:rtmodel}(a) Schematic representation
of the electron and hole bands crossing the Fermi level in the
superconducting state. (b) Interband recombination of QPs in the
$\alpha$ bands can occur with their counterparts in the electron
bands. As a result, a Cooper pair, an optical phonon and an
antiferromagnetic spin fluctuation are produced.}
\end{figure}

The simultaneous observation of nodal (i.e. $n_{th}\propto T^2$)
and fully gapped (i.e. $n\propto \Phi$) characteristics in our
data may be traced to the multiband nature of pnictides. ARPES
measurements have shown that five bands cross the Fermi surface:
three hole bands near zone center ($\Gamma$) and two at zone edge
(M) [Fig.~\ref{fig:rtmodel}(a)]. We attribute the entirety of the
observed signal to the three hole pockets at $\Gamma$, (i.e., the
inner hole bands $\alpha$ and outer hole band $\beta$, using the
nomenclature of \cite{ding}). Neither of the electron bands at M
($\gamma$ and $\delta$) contribute significantly to the signal as
LDA calculations \cite{ma} reveal a dearth of accessible states
1.5~eV above and (properly renormalized \cite{wray}) below the
Fermi level at this point in the Brillouin zone, rendering QP
recombination there optically dark at our probe wavelength.
Furthermore, an estimate of the Sommerfeld parameter shows that
the density of states at $\beta$ and at the combination of the two
bands at $\alpha$ are each 7 times larger than at $\gamma$ and
$\delta$ \cite{ding}, with the implication that the number of
photoinduced QPs in the hole pockets should overwhelm that of the
electron pockets.

The band structure and possible QP relaxation channels allow
assignment of the fast dynamics to the $\alpha$ band and the slow
component to $\beta$. The observation of a dispersion kink in
ARPES measurements at 25~meV that appears in $\alpha$ and
$\gamma$, $\delta$ but not $\beta$ \cite{richard,wray} indicates a
strong coupling between $\alpha$ and $\gamma$, $\delta$ by an
excitation at 13~meV, interpreted to be a resonant magnetic
excitation \cite{christianson}, and which arises due to their
near-perfect Fermi surface nesting. This excitation makes it
possible for the QPs in $\alpha$ to pair up with their
counterparts in $\gamma$, $\delta$ through an interband process by
emitting a combination of an optical phonon and a magnetic
excitation [Fig.~\ref{fig:rtmodel}(b)]. The inverse process,
however, is necessarily very slow since three-body scattering of
an antiferromagnetic spin fluctuation, an optical phonon and a
Cooper pair is required to generate QPs. The short lifetime of the
magnetic mode, as indicated by its relatively broad energy
linewidth \cite{christianson}, further inhibits the reverse
process, leading to a small value of $\gamma_{pc}$ which prevents
formation of a bottleneck in the interband channel. In the case of
intraband recombination in $\alpha$, which occurs by emission of
an $A_{1g}$ optical phonon ($E=2\Delta_0=24$~meV), its short
lifetime (3~ps)\cite{mansart} precludes it, too, from generating a
bottleneck. We therefore ascribe the intensity-dependent decay to
bare QP recombination in $\alpha$ and the slow dynamics to a boson
bottleneck in $\beta$. This bottleneck is enhanced by the
difference in gap energies between $\alpha$ ($2\Delta = 24$~meV)
and $\beta$ (2$\Delta = 12 $~meV), since, in addition to the
bosons already present from intraband recombination in $\beta$,
single bosons emitted from the interband recombination between
$\alpha$ and $\gamma$, $\delta$ may also break pairs in $\beta$.

These assignments allow for a proper interpretation of the
simultaneous observation of $n_{th}\propto T^2$ and $\Delta
R/R\propto \Phi$. The thermal QPs participating in the observed
recombination at low fluences are not necessarily the same as the
photoinduced ones; rather, they may also originate from either
$\gamma$ or $\delta$. A line node in either $\gamma$ or $\delta$
will lead to a preponderance of thermally excited QPs from the
electron bands as compared with the fully gapped hole bands. These
QPs will then dominate the recombination process which, in tandem
with the interband recombination process described above, can
account for the observed behavior in its entirety. While another
possibility for the quadratic temperature dependence of $n_{th}$
is the proposed $s_{\pm}$ gap symmetry with interband impurity
scattering, this scenario can only be realized within a very
specific range of impurity parameters \cite{chubukov}. Therefore,
while we cannot definitively differentiate between the two
scenarios at present, our results are highly suggestive of nodes
in the electron pockets.

In summary, we have presented time resolved measurements of
band-dependent QP dynamics in Ba$_{0.6}$K$_{0.4}$Fe$_2$As$_2$.
Below $T_c$, we find that the rate of QP recombination increases
linearly with excitation density indicating pairwise recombination
within the inner hole band. The number of photoinduced QPs
increases linearly with excitation density indicating fully gapped
nature of the hole bands. Meanwhile, the dynamics within the outer
hole band exhibit bottlenecked recombination. We have observed
that the thermal recombination rate of QPs varies as $~T^2$. This
likely arises from a line node in the electron bands although we
cannot definitively rule out the $s_{\pm}$ order parameter with
interband impurity scattering at this time.

The authors thank Dr. Deepak Singh for assistance with SQUID
magnetometry measurements, and Prof. B. Andrei Bernevig, Dr. David
Hsieh, and James McIver for useful discussions. This work was
supported by DOE Grant No. DE-FG02-08ER46521, NSFC, CAS and the
973 project of the MOST of China.

\bibliography{feas}

\end{document}